\documentstyle[aas2pp4]{article}

\begin{document}

\title{Variable Spin-down in the Soft Gamma Repeater SGR~1900+14 and
Correlations with Burst Activity}

\author{
Peter~M.~Woods\altaffilmark{1,3},
Chryssa~Kouveliotou\altaffilmark{2,3},
Jan~van~Paradijs\altaffilmark{1,4},
Mark~H.~Finger\altaffilmark{2,3},
Christopher~Thompson\altaffilmark{5},
Robert~C.~Duncan\altaffilmark{6}, 
Kevin~Hurley\altaffilmark{7},
Tod~Strohmayer\altaffilmark{8},
Jean~Swank\altaffilmark{8}, and
Toshio~Murakami\altaffilmark{9}
}

\altaffiltext{1}{Department of Physics, University of Alabama in Huntsville, 
Huntsville, AL 35899; peter.woods@msfc.nasa.gov}
\altaffiltext{2}{Universities Space Research Association}
\altaffiltext{3}{NASA Marshall Space Flight Center, SD50, Huntsville, AL
35812}
\altaffiltext{4}{Astronomical Institute ``Anton Pannekoek'', University of 
Amsterdam, 403 Kruislaan, 1098 SJ Amsterdam, NL}
\altaffiltext{5}{Department of Physics and Astronomy, University of North
Carolina, Philips Hall, Chapel Hill, NC 27599-3255}
\altaffiltext{6}{Department of Astronomy, University of Texas, RLM 15.308,
Austin, TX 78712-1083}
\altaffiltext{7}{University of California at Berkeley, Space Sciences
Laboratory, Berkeley, CA 94720-7450}
\altaffiltext{8}{NASA Goddard Space Flight Center, Greenbelt, MD 20771}
\altaffiltext{9}{Institute of Space and Astronautical Science, 3-1-1,
Yoshinodai, Sagamihara-shi, Kanagawa 229, Japan}

\begin{abstract}

We have analyzed {\it Rossi X-ray Timing Explorer} Proportional Counter Array
observations of the pulsed emission from SGR~$1900+14$ during September 1996,
June -- October 1998, and early 1999.  Using these measurements and results
reported elsewhere, we construct a period history of this source for 2.5
years.  We find significant deviations from a steady spin-down trend during
quiescence and the burst active interval.  Burst and Transient Source
Experiment observations of the burst emission are presented and correlations
between the burst activity and spin-down rate of SGR~$1900+14$ are discussed. 
We find an 80 day interval during the summer of 1998 when the average spin-down
rate is larger than the rate elsewhere by a factor $\sim$ 2.3.  This enhanced
spin-down may be the result of a discontinuous spin-down event or ``braking
glitch'' at the time of the giant flare on 27 August 1998.  Furthermore, we
find a large discrepancy between the pulsar period and average spin-down rate
in X-rays as compared to radio observations for December 1998 and January
1999.  

\end{abstract}


\section{Introduction}

Soft gamma repeaters (SGRs) form a rare class of persistent X-ray sources 
which are associated with young ($\sim$~10$^4$ year) supernova remnants (SNRs;
see Kouveliotou 1999a for a review).  Three of the four SGRs have stellar spin
periods within a narrow range of 5 -- 8 s (Mazets et al. 1979; Kouveliotou et
al. 1998; Hurley et al. 1999a); one SGR ($1627-41$) may rotate at 6.4 s (Woods
et al. 1999a), but the detection of this period is marginal and was not
confirmed in a recent ASCA observation (Hurley et al. 1999b). SGR~$1806-20$ and
SGR~$1900+14$ were recently found to spin-down on long time scales at a rate
$\sim$~10$^{-11}$ -- 10$^{-10}$ s s$^{-1}$ (Kouveliotou et al. 1998, 1999b). 
This spin-down has been interpreted as evidence that they are neutron stars
with very intense magnetic fields in the 10$^{14}$ -- 10$^{15}$ Gauss range,
i.e. magnetars (Duncan \& Thompson 1992).  Magnetars are defined as a star
whose magnetic field energy dominates all other sources of energy, including
rotation.  Except for their emitting brief ($\sim$~0.1 s), intense
($\sim$~10$^{39}$ -- 10$^{42}$ ergs s$^{-1}$) bursts of low-energy
$\gamma$-rays (Kouveliotou 1995) and having harder persistent emission spectra,
the characteristics of SGRs are similar to those of the anomalous X-ray pulsars
(AXPs; Mereghetti \& Stella 1995; van Paradijs, Taam, \& van den Heuvel 1995).

The spin-down histories of at least some SGRs and AXPs, have both a steady
spin-down component and a variable perturbing component.  Two AXPs with
well-sampled period histories, 1E~$1048.1-5937$ and 1E $2259+586$, have shown
evidence for such perturbations (Mereghetti 1995; Iwasawa, Koyama \& Halpern
1992; Heyl \& Hernquist 1998 and references therein).  SGR~$1806-20$ has a
long-term average spin-down rate of 8.3(3) $\times$ 10$^{-11}$ s s$^{-1}$,
although the local period derivative in November 1996 was 2.8(14) $\times$
10$^{-11}$ s s$^{-1}$ (Kouveliotou et al. 1998), suggesting a non-constant
spin-down.  

Recently, deviations from a constant spin-down rate have been found for
SGR~$1900+14$ (Kouveliotou et al. 1998; Woods et al. 1999b). SGR~$1900+14$ was
observed on 30 April 1998 with the Advanced Satellite for Cosmology and
Astrophysics (ASCA) when the source was not burst active, i.e. in quiescence. 
Timing analysis of its persistent X-ray flux revealed coherent pulsations with
a period of 5.16 s (Hurley et al. 1999a).  On 26 May 1998 the source became
extremely active (Hurley et al. 1999c), and has since remained in an active
state, during which numerous bursts have been recorded with  the Burst and
Transient Source Experiment (BATSE) aboard the {\it Compton Gamma-Ray
Observatory}.  Due to Earth occultation of the source for BATSE, the most
notable event from this source, emitted on 27 August 1998, went undetected. 
This exceptional flare (Hurley et al. 1999d) was much more energetic (factor of
$\sim$ 500) than the brightest burst emissions detected from this SGR before or
since, and rivals in intensity the brightest SGR outburst ever recorded, the
famous 5 March 1979 event from SGR~$0526-66$ (Mazets et al. 1979). 

We observed SGR~$1900+14$ with the {\it Rossi X-ray Timing Explorer} (RXTE)
Proportional Counter Array (PCA) at the beginning of its active period during
late May and early June 1998, as well as directly following the 27 August 1998
flare.  We confirmed the pulsar period (Kouveliotou et al. 1999b) and derived a
source spindown rate, $\sim$ 1 $\times$ 10$^{-10}$ s s$^{-1}$, hence
establishing SGR~$1900+14$ as a magnetar with $B_{\rm dipole}$ = 2 -- 8
$\times$ 10$^{14}$ G.  Comparing our data with the ASCA data, we noted the
spin-down of this magnetar was not constant from April through August, varying
from 5 -- 14 $\times$ 10$^{-11}$ s s$^{-1}$ (see Table 1 for period and period
derivative measurements).  Shitov (1999) reported the detection of radio
pulsations at 5.16 sec from SGR~$1900+14$ during December 1998 and January
1999 (see Table 1).

BeppoSAX Narrow Field Instrument (NFI) observations of SGR~$1900+14$ were
performed on 12 May 1997 and 15 September 1998.  The 12 May 1997 observation
provided a broad baseline for the spin-down during quiescence.  The average
spin-down rate between the BeppoSAX period measurement of May 1997 and the ASCA
measurement of April 1998 was 5.82(2) $\times$ 10$^{-11}$ s s$^{-1}$ (Woods et
al. 1999b).  This rate differs significantly from the values measured during
the burst active period, confirming that long-term variations in the spin-down
rate, possibly related to burst activity, occur.  An archival RXTE observation
from September 1996 extended the baseline of the quiescent spin-down, and
agrees to first order with the slower spin-down observed during quiescence
(Marsden, Rothschild \& Lingenfelter 1999).  

To further investigate the period history during the burst active period, we
have observed SGR~$1900+14$ with the RXTE PCA periodically between 3 January
1999 and 25 May 1999.  We combine a subset of these data with previously
reported results, in addition to a separate analysis of all RXTE observations
from September 1996 through October 1998 and construct a period history for
SGR~$1900+14$ over 2.5 years.  We also report on the burst rate history of
SGR~$1900+14$ as seen with BATSE and discuss possible correlations between the
burst activity and the changes observed in the spin-down.

\section{RXTE PCA Observations of Pulsed Emission}

SGR~$1900+14$ was observed with the RXTE PCA in September 1996 for 100 ksec
over 16 days.  An analysis of a subset of these data provided a period and
period derivative of 5.1558199(29) s and 6.0(10) $\times$ 10$^{-11}$ s
s$^{-1}$, respectively at the chosen epoch 50338.216 MJD (Marsden et al.
1999).  In order to compile a uniform database, we have re-analyzed these data
using a phase folding technique.  The first 82 ksec were acquired over 12 days
with a pointing offset of 0.01$^{\circ}$.  Using event mode data, we energy
selected these observations for 2 -- 10 keV photons, binned the data at 0.125 s
time resolution and barycenter corrected the bin times.  Adopting a constant
period phase model we derived from an epoch fold search, we calculated the
phase at multiple points during the observation.  These phases could not be
well fit with a linear phase model, so we included a second order term (i.e.
frequency derivative).  This fit yielded a good reduced $\chi^2$ of 0.8 for 5
degrees of freedom (see Table 1).  The phase-folded profile (2 -- 10 keV) for
this ephemeris is given in Figure 1a.  The period and period derivative
derived from this fit (see Table 1) are similar to, but not within the errors
reported by Marsden et al. (1999).  We estimate the average spin-down rate
between September 1996 and April 1998 (quiescence) by performing a least
squares fit to all period measurements during this time interval and find a
value 6.13(2) $\times$ 10$^{-11}$ s s$^{-1}$.  The statistical error here has
been inflated by the square root of the reduced $\chi^2$ (168) of the linear
fit.  The period derivative measurement for September 1996 has been
extrapolated for 250 days in Figure 2 (dotted lines represent $\pm 1\sigma$) to
clearly indicate the discrepancy between the local slope and the long-term
trend.  The improved precision of the period derivative measurement within this
observation and the inclusion of the BeppoSAX period measured from May 1997
allows us to conclude that there are significant deviations from a constant
spin-down rate during quiescence.

We have also re-analyzed RXTE PCA observations of SGR~$1900+14$ during May/June
1998 and August-October 1998.  For the first sequence of observations from 31
May through 9 June 1998, we processed the data as before and find a second
order phase model represents the data well with a reduced $\chi^2$ of 1.3 for
13 dof (see Table 1).  The phase folded profile (Figure 1b) is consistent with
the lightcurve found in September 1996, which shows the profile did not change
over a long time period (1.7 years), as well as after significant burst
activity.  We have extended the baseline of the second sequence of observations
from 6 days (Kouveliotou et al. 1999b) to 42 days.  Due to the large number of
bursts detected during these observations (more than 1000), we have binned the
data with finer time resolution (0.05 sec) to better filter out these bursts. 
Over this much longer baseline, a simple second-order phase model fits the data
reasonably well, but not completely ($\chi^2$ = 2.4 for 25 dof).  The phase
folded profile (Figure 1c) is significantly different during these observations
as pointed out by Kouveliotou et al. (1999b) and shows the pulse shape changes
observed during the tail of the August 27$^{\rm th}$ flare (Mazets et al. 1999)
persist for months.  Combination of these observations with a more recent RXTE
observation campaign will allow us to better investigate these low phase
amplitude deviations.

A new series of RXTE PCA observations of SGR~$1900+14$ began on 3 January 1999
and are currently on-going through 25 May 1999.  We have selected a subset of
three observations sufficiently long to obtain an accurate measurement of the
pulsar period.  For each of these observations, we have used Standard 1 data (2
-- 60 keV) binned at 0.125 s time resolution with barycenter corrected time
bins.  Using a phase folding technique, we have calculated the period for each
observation.  The results of these measurements are summarized in Table 1. 
Although summed over a different energy range, the phase folded profile is the
same as found during the fall of 1998.  The X-ray periods measured during 1999
lie slightly above the extrapolation of the period derivative found during the
fall of 1998.  We find a linear least squares fit to the period measurements
after 27 August 1998 yields an average period derivative 6.07(15) $\times$
10$^{-11}$ s s$^{-1}$; again, the statistical error is inflated by the square
root of the reduced $\chi^2$ (56).  We note that the reported radio period
measurement during December 1998 and January 1999 is highly discrepant (the
reported period is more than 5000$\sigma$ away from the linear fit, and the
period derivative is double what is found in X-rays over the same time
interval; see Figure 2).

\placetable{tbl-1}

\begin{noindent}
\begin{center}
\begin{deluxetable}{cccllcc}
\footnotesize
\tablecaption{Period and Period Derivative Measurements for SGR~$1900+14$ 
\label{tbl-1}}
\tablewidth{0pt}

\tablehead{
\colhead{Time of}               &
\colhead{Exposure}              &
\colhead{Epoch}                 &  
\colhead{Period}                &  
\colhead{Period}                &  
\colhead{Instrument}            &  
\colhead{Reference}             \\
\colhead{Observation}           &
\colhead{}                      &
\colhead{}                      &
\colhead{}                      &
\colhead{Derivative}            &
\colhead{}                      &
\colhead{}                      \\
\colhead{mm/dd/yy}              &
\colhead{ksec}                  &
\colhead{MJD TDB}               &
\colhead{s}                     &
\colhead{10$^{-11}$ s s$^{-1}$} &
\colhead{}                      &
\colhead{}
}

\startdata

09/04/96 - 09/16/96  &   82.1  &  50337.0  &  5.1558157(3)   &  8.42(23)     &
RXTE      &  This work  \nl 

05/12/97 - 05/13/97  &   45.7  &  50580.5  &  5.157190(7)    &  $--$         &
BeppoSAX  &  Woods et al. 1999b  \nl 

04/30/98 - 05/02/98  &   84.6  &  50935.0  &  5.1589715(8)   &  $--$         &
ASCA      &  Hurley et al. 1999a  \nl 

05/31/98 - 06/09/98  &   43.5  &  50970.0  &  5.15917011(55) &  8.2(6)       &
RXTE      &  This work  \nl 

08/28/98 - 10/08/98  &  146.9  &  51070.0  &  5.16026572(12) &  5.93(3)      &
RXTE      &  This work  \nl 

09/15/98 - 09/16/98  &   33.2  &  51071.5  &  5.160262(11)   &  $--$         &
BeppoSAX  &  Woods et al. 1999b  \nl 

09/16/98 - 09/17/98  &   39.0  &  51073.3  &  5.160295(3)    &  $--$         &
ASCA      &  Murakami et al. 1999  \nl 

12/12/98 - 02/04/99  &   $--$  &  51159.5  &  5.16129785(8)  &  12.3228(34)  &
BSA       &  Shitov 1999  \nl 

01/03/99 - 01/04/99  &   31.9  &  51181.5  &  5.160934(56)   &  $--$         &
RXTE      &  This work  \nl 

03/21/99 - 03/21/99  &    9.1  &  51259.0  &  5.16145(18)    &  $--$         &
RXTE      &  This work  \nl 

03/30/99 - 03/30/99  &    8.4  &  51268.0  &  5.16156(11)    &  $--$         &
RXTE      &  This work  \nl

\enddata

\end{deluxetable}
\end{center}
\end{noindent}

\section{BATSE Observations of Burst Emission}

Between May 1998 and January 1999, BATSE triggered on 63 bursts from
SGR~$1900+14$.  The on-board BATSE trigger criteria were set at low-energy
trigger (Channels 1 + 2; 25 -- 100 keV), nominal trigger (Channels 2 + 3; 50 --
300 keV) and high-energy trigger (Channels 3 + 4; 100 -- 2000 keV), each for
significant time intervals during the burst active period.  Due to the
relatively soft nature of the typical burst emission from SGRs, BATSE's
sensitivity to SGR events changed according to which trigger criterion was in
use.  Other factors, such as inability to read out on-board memory before the
next event, resulted in untriggered events as well.  In order to obtain a more
complete database, we performed an off-line search for untriggered events from
SGR~$1900+14$.

BATSE consists of eight NaI detectors which form a regular octahedron and are
sensitive to photons with energies 25 keV -- 2 MeV (Fishman et al. 1989). 
Using CGRO spacecraft pointing information, we calculated the zenith angles for
each detector and determined the two detectors with the lowest zenith angles
for each spacecraft orientation.  For all days during each orientation, we
searched the DISCLA data (1.024 s time resolution) for simultaneous
fluctuations in these detectors for energy channel 1 (25 -- 50 keV).  The
background was estimated by fitting a first-order polynomial to 10 seconds of
data before and after each bin with a 3 second gap between the background
interval and the bin searched.  An off-line trigger was defined as a
fluctuation greater than 4.5 $\sigma$ and 3.0 $\sigma$ in the two brightest
detectors, respectively, an excess of counts below 50 keV, relative to the
counts between 50 and 300 keV for the brightest detector and a duration less
than 7 seconds.  Each trigger was visually inspected and coarsely located based
upon the relative rates in the BATSE modules.  Between 24 May 1998 and 3
February 1999, we detected most of the SGR~$1900+14$ events which triggered
BATSE in addition to 137 untriggered events.  Some triggered events were not
detected off-line because the DISCLA data has coarser time resolution than the
time scale on which most SGR events trigger (64 msec) or the trigger occurred
during a telemetry data gap.  The large number of untriggered events relative
to triggered bursts is due to extended periods (more than 4 months) when the
BATSE trigger was in `high-energy mode.'  Based upon our experience with
classifying BATSE triggers, we expect the number of false triggers within our
sample to be less than 5\%.  Figure 2 displays the burst rate (per 3 day
interval) over the time period searched.  No emission from SGR~$1900+14$
triggered the BATSE instrument between August 1992 (Kouveliotou et al. 1993)
and 25 May 1998.  We note that the burst activity is currently low, but has not
ceased as of the writing of this manuscript.

\section{Discussion}

We have shown that during quiescence, the spin-down rate of SGR~$1900+14$ is
not constant.  The deviations observed may be caused by processes such as:
orbital Doppler shifts; persistent but variable emission of Alfv\'en waves
and/or particles from magnetars (Thompson \& Blaes 1998); radiative precession
in such an object (Melatos 1999); or discontinuous spin-up events (glitches) as
seen in radio pulsars (Thompson \& Duncan 1996; Heyl \& Hernquist 1998).  Due
to the sparsely sampled data for this source, we cannot exclude any of these
models currently.  More frequent measurements are required before anything
definitive can be said about modeling these deviations.

The period history of SGR~$1900+14$ appears to be divided into two sections
during which the spin-down rate is nearly, but not exactly, constant.  The
pulse period and period derivative reported by Shitov (1999) for a radio pulsar
connected with SGR~$1900+14$ appears incompatible with the X-ray spin history
presented here.  Before 9 June 1998 and after 27 August 1998, the average rate
is 6.1 $\times$ 10$^{-11}$ s s$^{-1}$.  These two sections are separated by 80
days during which the period increased by 1 ms, which implies an average rate
of 14.0 $\times$ 10$^{-11}$ s s$^{-1}$.  It appears that the period history
during this interval allows for two obvious descriptions: (i) a gradual
increase of the nominal spin-down rate and (ii) a discontinuous spin-down event
associated with the 27 August 1998 flare.

According to the first picture, following the initial flurry of burst activity
in late May 1998, the spin-down rate increased by a factor $\sim$ 2.3.  This
rate persisted for $\sim$ 80 days, then decreased to near its original value
after 27 August 1998.  If this were the correct description, then we cannot
attribute the enhanced spin-down directly to the magnitude of the burst
activity.  The number of bursts recorded with BATSE between the onset of
activity in May and 26 August 1998 was 40 (the extraordinary, multi-episodic
burst of 30 May 1998, Hurley et al. 1999c, is counted here as a single event).
Following the 27 August 1998 flare, 123 bursts were recorded up to 3 February
1999, including a multi-episodic event on 1 September 1998 similar to, although
less intense than the 30 May 1998 event. The total burst energy recorded in the
events following the 27 August flare is more than double the energy released
through burst emission prior to the flare.  If one assumes that the burst rate
or the burst energy released is correlated with the increase in spin-down, then
one would expect to see an even steeper spin-down rate between 28 Aug 1998 and
3 January 1999, which is not the case.  The increased breaking torque on the
star must then be attributed to something other than the burst activity as
measured by the burst rate or the burst energy of the smaller, more common
bursts.  

An alternative scenario to account for the rapid spin-down during the period
June -- September 1998 is that the star underwent a more or less steady
spin-down at a long-term average rate of $\sim$ 6 $\times$ 10$^{-11}$ s
s$^{-1}$ from June through most of August.  The star then suffered a
discontinuous upward jump in period or a ``braking glitch'', which is
attractive to link with the occurrence of the very energetic flare of 27 August
1998.  Extrapolating the long-term trends found before and after 27 August, we
find that this braking glitch would have a magnitude $\Delta$P = 5.72(14)
$\times$ 10$^{-4}$ s.  This corresponds to a rotational energy loss for the
star $\Delta$E$_{\rm rot} \approx 2\times 10^{41} (M_\star/1.4
M_\odot)\,(R_\star/10~{\rm km})^2$ ergs if the whole star participates, or
0.5\% of the energy released in high-energy photons during the 27 August
flare.  Physical mechanisms causing such a glitch are discussed in a companion
paper (Thompson et al. 1999).

\acknowledgments{\noindent {\it Acknowledgements} -- We thank the RXTE SDC for
pre-processing the RXTE data.  PMW acknowledges support under the cooperative
agreement NCC 8-65.  JvP acknowledges support under NASA grants NAG 5-3674 and
NAG 5-7060.}

\newpage

\figcaption{Phase folded profiles of SGR~$1900+14$ as seen with the RXTE PCA (2
-- 10 keV) for (a) September 1996, (b) May -- June 1998, and (c) August --
October 1998.}

\figcaption{{\it Bottom} -- Period history of SGR~$1900+14$ from September 1996
through March 1999.  Lower axis label is modified julian date and upper axis is
mm/dd/yy.  The solid lines indicate least square fits to the period
measurements found in two separate intervals (September 1996 -- June 1998 and
August 1998 -- March 1999).  Due to the long series of observations with RXTE
from 28 August to 8 October 1998, two period measurements from the beginning
and end are shown.  Residuals of fit are shown in lower panel.  Dotted lines
represent extrapolation of local period derivative measurement ($\pm 1\sigma$;
see Table 1) found in September 1996 RXTE observation.  {\it Right} -- Inset of
lower figure showing burst rate history (upper panel) and period history (lower
panel) of SGR~$1900+14$ from 7 April 1998 through 16 February 1999.  Dotted
lines represent extrapolation of local period derivative measurement ($\pm
1\sigma$).}

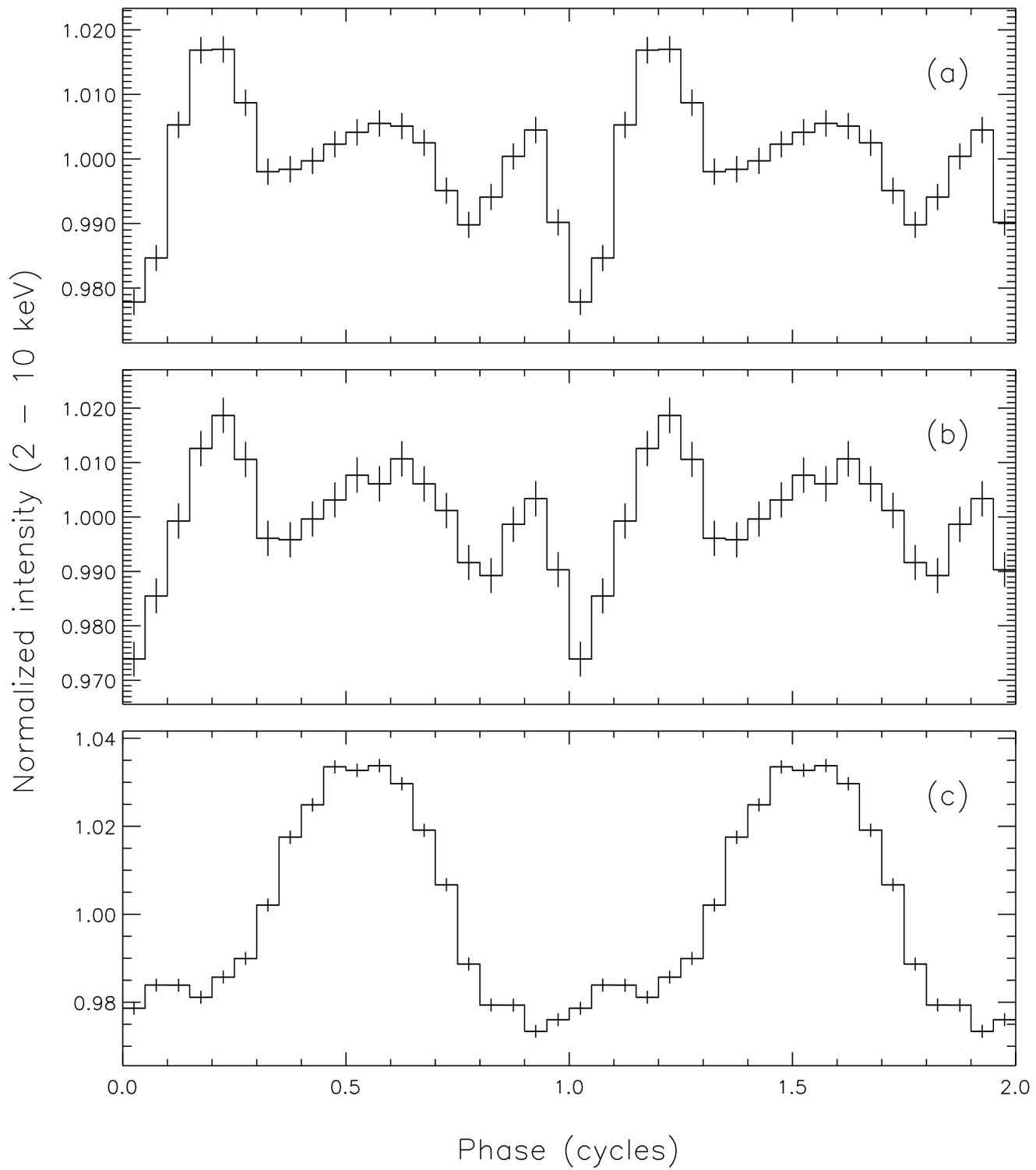
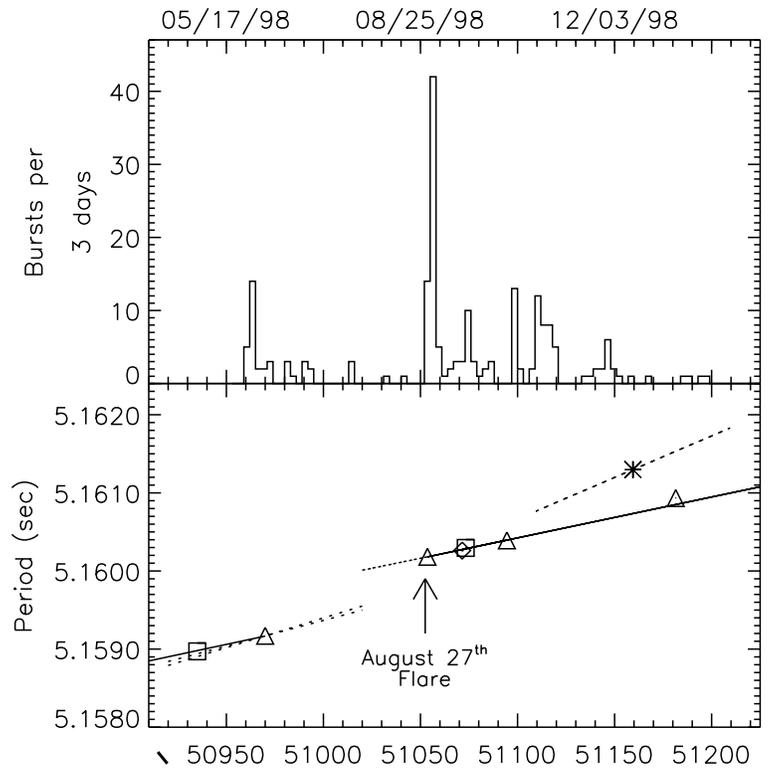
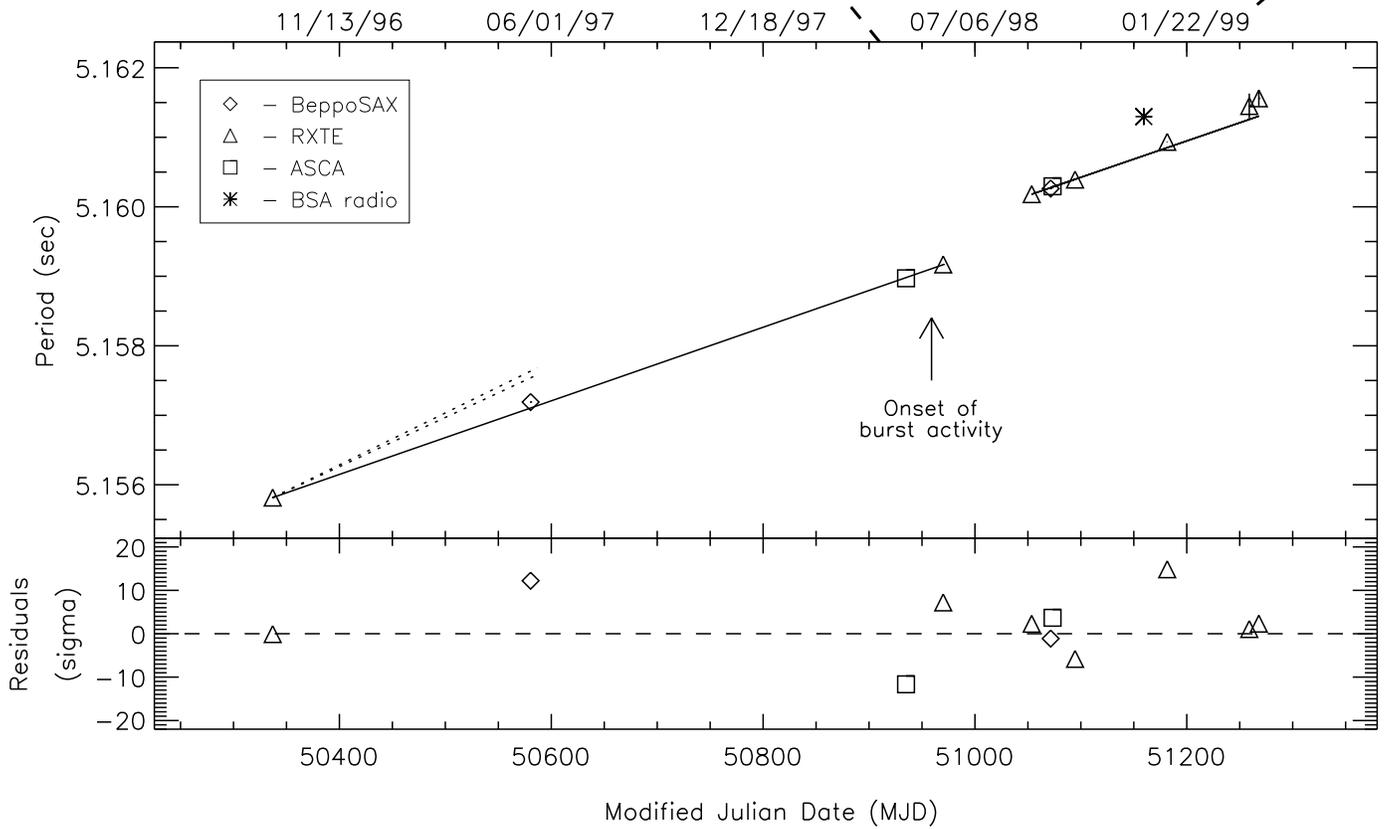
\end{document}